\documentclass[aip,apl,reprint,superscriptaddress]{revtex4-1}
\usepackage{graphicx}
\usepackage{color}
\usepackage{epstopdf}
\begin{document}


\title{High resolution structural characterisation of laser-induced defect clusters inside diamond}

\author{Patrick S. Salter}
\affiliation{Department of Engineering Science, University of Oxford, Parks Road, Oxford, OX1 3PJ, UK}
\author{Martin J. Booth}
\affiliation{Department of Engineering Science, University of Oxford, Parks Road, Oxford, OX1 3PJ, UK}
\author{Arnaud Courvoisier}
\affiliation{Department of Engineering Science, University of Oxford, Parks Road, Oxford, OX1 3PJ, UK}
\author{David A.J. Moran}
\affiliation{School of Engineering, University of Glasgow, Glasgow G12 8QQ, UK}
\author{Donald A. MacLaren}
\affiliation{SUPA, School of Physics \& Astronomy, University of Glasgow, Glasgow G12 8QQ, UK}
\email{patrick.salter@eng.ox.ac.uk}
\email{damaclaren@physics.org}

\date{\today}

\begin{abstract}
Laser writing with ultrashort pulses  provides a potential route for the manufacture of three-dimensional wires, waveguides and defects within diamond. We present a transmission electron microscopy (TEM) study of the intrinsic structure of the laser modifications and reveal a complex distribution of defects. Electron energy loss spectroscopy (EELS) indicates that the majority of the irradiated region remains as $sp^3$ bonded diamond. Electrically-conductive paths are attributed to the formation of multiple nano-scale, $sp^2$-bonded graphitic wires and a network of strain-relieving micro-cracks. 
\end{abstract}

\maketitle

Diamond has long been exploited for its extreme mechanical properties but advances in the synthesis of large single crystals also make it attractive for a variety of new technologies, including harsh environment applications~\cite{RD42_1999}, high frequency electronics~\cite{Russell2012}, biosensors~\cite{Picollo2013} and photonics~\cite{Aharonovich2011}. In particular, the recent use of nitrogen vacancy (NV) centers for both quantum processing~\cite{Bernien2013} and as extremely sensitive magnetic field and temperature sensors~\cite{Balsub2008, Maze2008, Neumann2013, clevenson2015} has stimulated intense research.

Direct laser writing (DLW) with ultrashort pulses is a versatile, emerging platform for diamond functionalisation that could benefit all of these applications. Sub-micrometre scale features can be processed over large volumes in manageable timescales and there are new prospects for three dimensional device architectures fabricated directly within the bulk material. Aside from surface modification~\cite{Shinoda2009, Lehmann2014, Komlenok}, there are two key regimes for sub-surface processing: (i) at very low pulse energy the highly non-linear interaction generates an ensemble of vacancies at the laser focus~\cite{Chen2017} while (ii) at higher pulse energies, there is break-down of the diamond lattice leaving a conductive graphitic phase~\cite{kononenko2008}. Regime (i) is an important precursor for the formation of coherent NV centers for quantum applications~\cite{Chen2017, Hadden2017}. Regime (ii) enables the creation of embedded electrodes, particularly for use in 3D radiation detectors offering a superior radiation tolerance and a faster response than their planar counterparts~\cite{Caylar2013a, Oh2013b, Lagomarsino2013a}. Furthermore, Regime (ii) modifications provide a potential route for all-carbon metamaterials~\cite{Shimizu2009}, solar cells~\cite{Trucchi2017}, photonic crystals~\cite{Kononenko2014} and waveguides~\cite{Courvoisier2016, Sotillo2016}. For successful development of DLW in these applications and other diamond based  technologies, it is of vital importance to carry out detailed structural analysis of the subsurface laser induced changes to the diamond. 

\begin{figure}[t]
\begin{center}
\includegraphics[width=8.5cm]{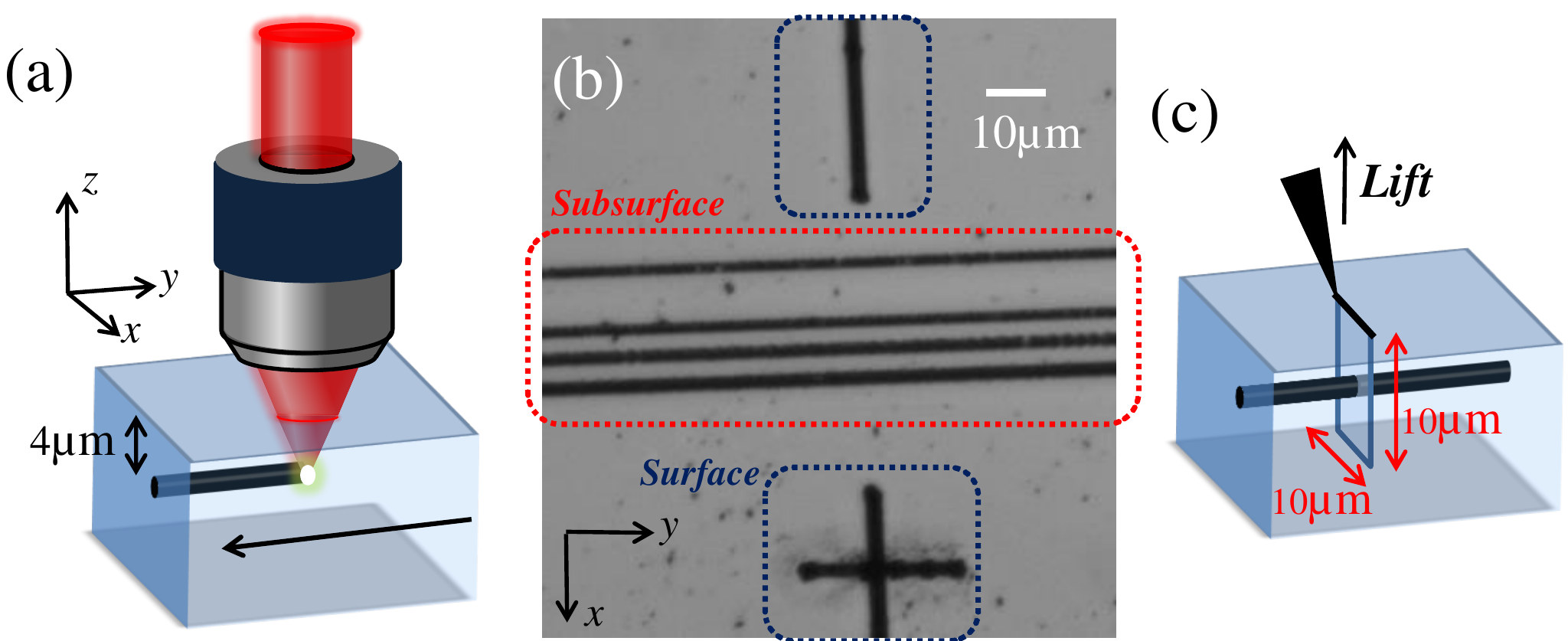}
\caption{\small{\emph{(a) Graphitic wires were laser written parallel to the top surface of the diamond at a depth of $4~\mu$m. (b) Optical transmission microscope image of the four subsurface wires (centre) with registration marks laser written onto the surface to either side. (c) A FIB was used to mill a cross-section from the wires, which was lifted out for TEM analysis.  }}
\label{Fig1}}
\end{center}
\end{figure}

When imaged in an optical microscope, the structural modifications in Regime (ii) appear uniform and strongly absorbing, suggesting a bulk conversion of diamond into graphite. However, previous Raman studies revealed only the partial formation of \textit{sp$^2$} bonded graphite within the laser irradiated zones~\cite{kononenko2008, Shimizu2009, SalterSPIE2014, Sotillo2016}. When using such modifications to create wires, the resultant resistivity varies from $0.02 - 2~\Omega$cm~\cite{Lagomarsino2013a, Sun2014b, Lagomarsino2014d, Shimizu2009, Trucchi2017}, with values at least an order of magnitude higher than that for polycrystalline graphite~\cite{J.D.Cutnell2004a}. Similarly, when writing optical waveguides using the stress field generated by modifications in Regime (ii), the propagation losses are significantly lower than those expected from a complete conversion of the irradiated zones to graphite~\cite{Sotillo2016}. These observations indicate that the wires' internal structure is more complex than suggested by optical imaging. As illustration, a recent scanning electron microscopy (SEM) study of subsurface DLW structures exposed by mechanical polishing revealed a main fracture containing graphenic carbon and smaller conductive nanocracks propagating from it~\cite{kononenko2016}. Here, we use transmission electron microscopy (TEM) to provide depth resolution and better spatial resolution of these sub-surface structures and employ electron energy loss spectroscopy (EELS) to assess composition. The results provide a vital perspective on the nature of DLW subsurface modifications in diamond, advancing our understanding of both processing regimes.

The diamond was a single crystal CVD sample (Element 6 optical grade, 3~mm square) cut with $\left\{100\right\}$ edges. A regeneratively amplified titanium sapphire laser producing a 1~kHz train of 250~fs, $25$~nJ pulses at a wavelength of 790~nm was focused $4~\mu$m beneath the surface of the diamond using an oil immersion objective lens with a numerical aperture of 1.4. A liquid crystal spatial light modulator was used to correct for focal distortion arising from refraction at the diamond surface, but we note that the aberration is relatively small at such a shallow depth. The diamond was translated through the laser focus at $10~\mu$m/s to produce subsurface tracks parallel to the diamond's top surface, as illustrated in Fig~\ref{Fig1}(a); further details can be found elsewhere~\cite{Sun2014b}. Wires were thus fabricated and the sample was annealed at $900^{\circ}$C in a nitrogen atmosphere for one hour. 

A subsurface group of four wires can be seen in the optical transmission microscope image in Fig.~\ref{Fig1}(b), along with laser-written surface registration marks. The wires appear optically to have a width around $2\mu$m, with edge roughness on a 100~nm length-scale.  The resistivity was measured for the wires as 1.6~$\Omega$~cm, which is toward the top of the range reported in the literature. The relatively high resistivity can be understood by noting the difficulties in writing buried graphitic wires both parallel and close to the top diamond surface. The pulse energy has to be low to avoid any surface damage and a high degree of axial confinement is necessary for the subsequent processing,  precluding the use of an axial multi-scan fabrication technique to improve the wire conductivity~\cite{Sun2014b, Trucchi2017}.

Cross-sections of wires were extracted by using standard `lift-out' protocols (e.g. see Ref. \cite{FIBLiftOut}) on an FEI Nova DualBeam Focused Ion Beam (FIB) microscope. A protective Pt cap was deposited prior to ion milling and a typically $(10 \mu\mathrm{m})^2$ cross-section through the wire was removed and ion-polished to $<150$nm thickness (i.e. thicker than usually desired for high resolution TEM because the final low-energy FIB polishing step was less effective than for softer materials). TEM and scanning TEM (STEM) were performed on a JEOL ARM cFEG instrument operated at 200~kV, using a Gatan Quantum spectrometer for electron energy loss spectroscopy (EELS). The EELS data were acquired using the `spectrum imaging' methodology \cite{SpectrumImaging} and processed using standard routines for background subtraction and deconvolution within Gatan's Digital Micrograph software. 

\begin{figure}[t]
\begin{center}
\includegraphics[width=8.5cm]{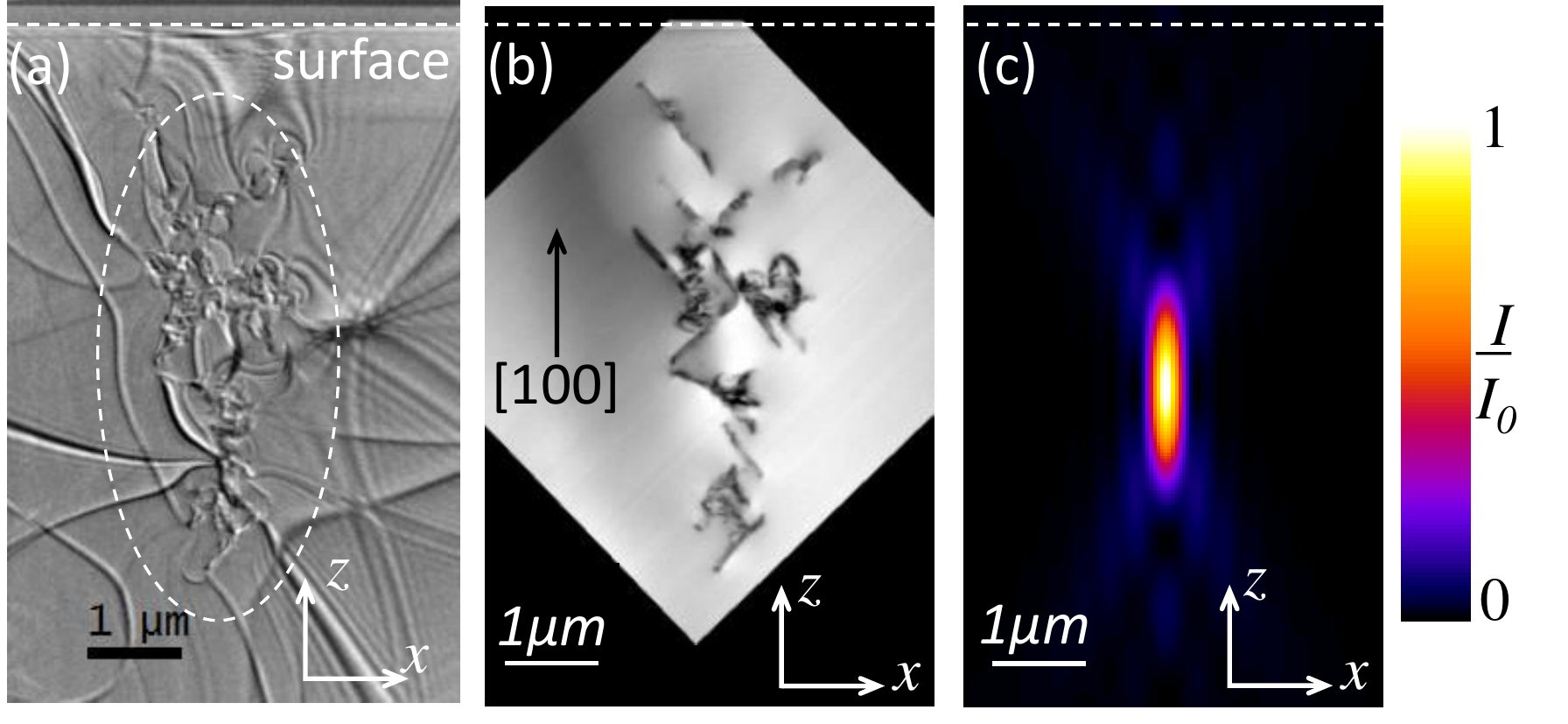}
\caption{\label{Fig2}\small{\emph{(a) Cross-sectional TEM image of an annealed wire. Contrast is dominated by electron interference but the dashed ellipse indicates the fine structures caused by laser treatment. (b) Bright field STEM image of the same area, without interference effects and revealing a cluster of discrete structural modifications. (c) The calculated laser intensity profile (to scale). The dotted horizontal line indicates the diamond surface.}}}

\end{center}
\end{figure}

The extent of sub-surface modifications is revealed by TEM in Fig.~\ref{Fig2}, which plots both (a) TEM and (b) bright-field STEM images of an annealed wire cross-section. Contrast in the TEM image is dominated by interference effects due to variations in the thickness and angle of the sample, producing radial black and white lines from strain within the modified region. Fine details within the annotated ellipse are attributed to DLW structural modifications. Interference effects are largely absent from the STEM image, which picks out more clearly disruptions of the diamond lattice as dark features. Visible damage is constrained within a region that is 1.9 $\mu$m wide, lying between 0.5 $\mu$m and 5.0 $\mu$m beneath the diamond surface. Although the elongated shape of the damage region matches that of the laser intensity profile, it is interesting that its location is not clearly centred on the laser focus, which is illustrated to scale in Fig.~\ref{Fig2}(c) and centred at a depth of 4 $\mu$m. 

The most striking observation of Fig.~\ref{Fig2}(b) is that there is no single homogenous region of modified diamond, rather the wires comprise a disperse cluster of smaller structures. Their irregularity is consistent with the slightly ragged wire edges observed optically. The majority of them are individually less than a few hundred nanometres in size, so below the optical diffraction limit, but sufficiently closely spaced that the wire appears as a single continuous feature when imaged optically. We attribute the optical contrast [see Fig~\ref{Fig1}(b)] to a combination of absorption, scattering from defects and refractive index changes arising from strain fields  that propagate out from the defects. This demonstrates that standard optical methods lack the resolution to accurately characterize DLW wires in diamond. 

\begin{figure*}[t]
\begin{center}
\includegraphics[width=13cm]{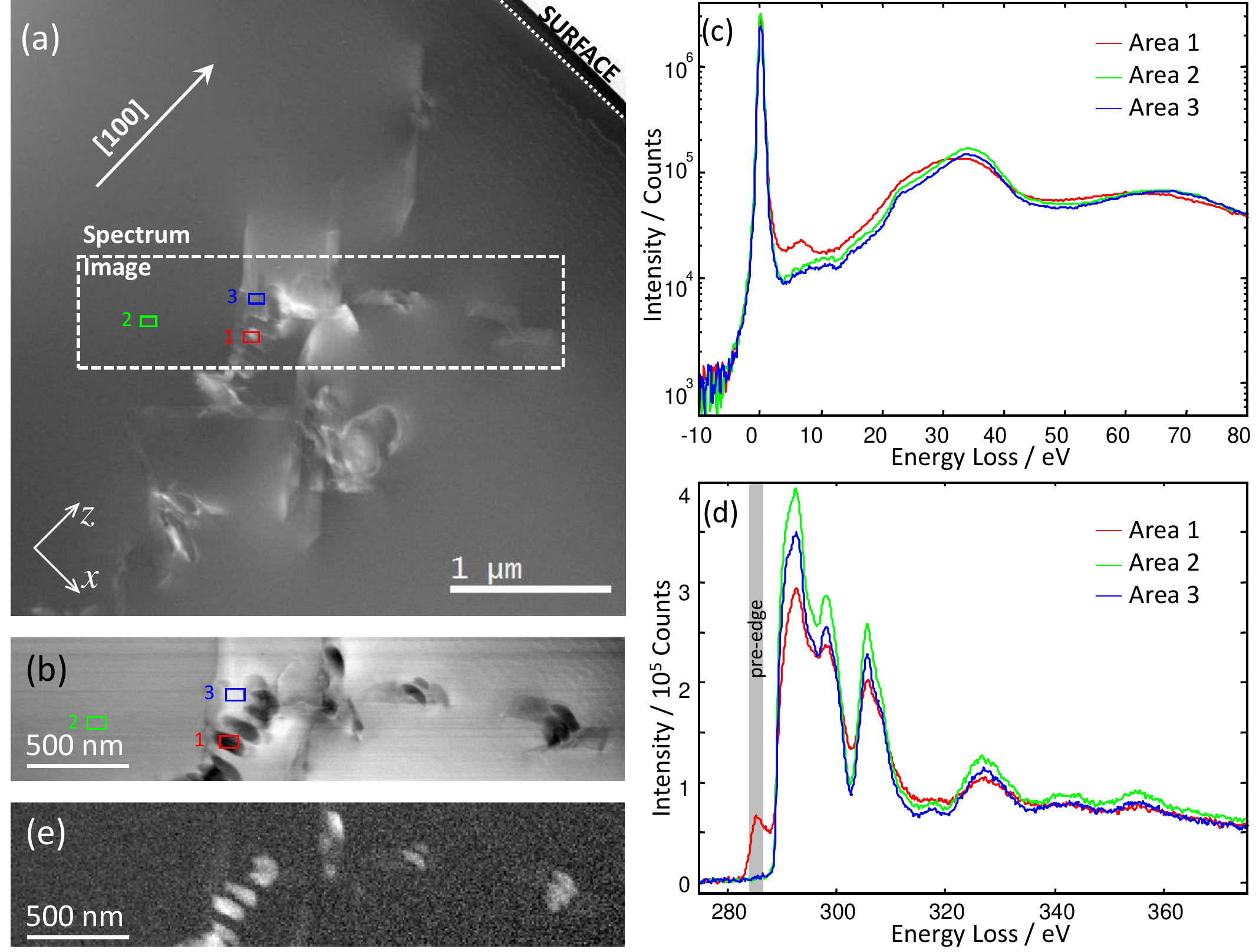} 
\caption{\label{Fig3}\small{\emph{EELS characterisation of an annealed wire. (a) Dark-field STEM image of the same region illustrated in Fig. \ref{Fig2} (rotated by $45^\circ$). The [100] direction is indicated. (b) Low-loss and (c) processed core-loss EELS spectra collected from three regions indicated in the STEM image. (d) The high-angle dark field image collected within the region labelled `spectrum image' and (e) the spatial distribution of the pre-edge ($sp^2$bonding) feature indicated in the core-loss spectra within the spectrum image region.}}}
\end{center}
\end{figure*}

The bonding configuration and chemistry of the defects are revealed by EELS, which is summarised for an annealed wire in Fig.~\ref{Fig3}. A $3~ \times 0.7\mu\mathrm{m}^2$ region straddling the structural features was used for spectrum imaging, where both low-loss and core-loss EELS spectra were acquired on a pixel-by-pixel manner. Note that Fig.~\ref{Fig3}(a) is a dark field image and so has reversed contrast compared to Fig.~\ref{Fig2}(b). Fig.~\ref{Fig3}(b), on the other hand, is a high angle annular dark field (HAADF) image with contrast that derives principally from mass or thickness variations, with denser regions appearing brightest. Together, the two images reveal the structural modifications as a combination of (i) discrete oval patches that appear dark in HAADF and so are less dense than the surrounding matrix and (ii) sharp diagonal lines lying at $45^\circ$ to the [100] direction. 

Typical EELS low-loss and carbon K-edge spectra are shown in Fig.~\ref{Fig3}(c) and (d), respectively, and were acquired from the areas indicated in Fig.~\ref{Fig3}(a). Area 1 is located within one of the oval patches; area 2 lies within a nearby bright feature; and area 3 represents the unmodified diamond. Both the core-loss and the low-loss spectra reveal area 1 to differ chemically and electronically from the surrounding material. 

The diamond low-loss spectrum (area 3) is dominated by an asymmetric peak with a maximum at 34.4~eV that agrees with previous studies of the diamond bulk plasmon ($\approx$35~eV) with an asymmetry caused by a surface excitation around 25~eV~\cite{CanMineral33_1157, JNucMat191_346, PhilMag86_4757}. Area 2's spectrum is similar while the plasmon in area 1 shifts downwards in energy, an effect that has previously been attributed to damage~\cite{JNucMat191_346} and reflects a reduction in valence electron density~\cite{PhilMag87_4073}. Amorphous carbon, carbon onions, fullerenes and graphite are all known to have bulk plasmons at lower energies (around 24, 24.5, 26 and 27~eV respectively~\cite{Carbon49_5049}) and it would be difficult to decompose the spectrum here into distinct components. The sharp peak at 7~eV, however, is more distinctive. It is at too high an energy for purely amorphous carbon~\cite{NanoLett9_1058} but has previously been attributed to the $\pi^*$ plasmon excitation of graphite~\cite{NanoLett9_1058, NatureNano3_676} and has also been observed from `brown diamond', which is known to contain intrinsic defects of $sp^2$ character \cite{PhilMag86_4757}.

Turning to the spectra in Fig.~\ref{Fig3}(d), previous K-edge studies have shown clear spectral differences between carbonaceous materials~\cite{CanMineral33_1157}. Pure diamond typically shows a sharp exciton around 289~eV, followed by multiple overlapping $\sigma^*$ excitations at higher energies~\cite{APL69_568}, as observed from areas 2 and 3. The pronounced dip in intensity 302.5~eV is caused by a bandgap in the unoccupied density of states (and is therefore a good indication of insulating or semiconducting character)~\cite{APL69_568} whilst a subsequent peak at 327~eV is caused by multiple scattering events and is a good indication of crystallographic order~\cite{PRB76_094201}. Similar to previous studies~\cite{Porro}, area 1 exhibits an additional weak `pre-edge' peak at 285.5~eV that is attributed to excitation to a $\pi^*$ state and is therefore indicative of the formation of unsaturated, $sp^2$ bonded carbon. The spectrum from area 1, however, retains the other distinctive diamond features (notably the dip and peak at 302.5~eV and 327~eV, respectively) and lacks either the broad, featureless profile above 289~eV that is typically seen in amorphous or defective carbon, or the sharp exciton and $\sigma^*$ features that are shifted above 291~eV for graphite~\cite{CanMineral33_1157}. Both low-loss and K-edge spectra from area 1 are  therefore consistent with the coexistence of $sp^3$-bonded, diamond-like material and $<20\%$ of $sp^2$ bonded material. The weakened band-gap dip and the emergence of peaks at both 7~eV and 285.5~eV are all consistent with the existence of electrically-conductive carbon phases and therefore underpin the formation of conductive wires by DLW. Area 2, which lies 200~nm away from a conductive region, shows no evidence of $sp^2$ carbon, illustrating the extremely compact nature of the conductive regions.

The spatial distribution of the mixed bonding regions is shown in Fig.~\ref{Fig3}(e) , which maps the strength of the `pre-edge' $\pi^*$ feature and shows a clear correlation with the dark oval features observed by HAADF. Together, the data indicate that conversion of diamond to graphitic carbon occurs in a series of discrete oval patches that lie along the surface normal direction and are $200$nm $\times 80$~nm in size. The diagonal features from Fig.~\ref{Fig2}(b) (oriented at $\approx 45^\circ$ to the surface normal) are barely visible in Fig.~\ref{Fig3}(e), but do still produce a weak pre-edge component, suggesting that they too may be electrically conductive. These features lie preferentially  along the $<110>$ directions, and are therefore likely contained in $\left\{111\right\}$ cleavage planes. Each set is accompanied by an $sp^2$ patch, including those closest to the diamond surface and furthest from the laser focus. This strongly suggests that these diagonal features are strain-relieving dislocations that form as a consequence of the volume increase on transformation of the diamond. Also of note is an apparent periodicity in the formation of the $sp^2$ clusters, such as the sequence of oval patches in the upper middle of Fig.~\ref{Fig3}(e). All wires analyzed displayed a similar ordering with an approximate periodicity of 110~nm along the direction of propagation for the laser, suggesting that their formation is not random. Indeed, it is similar to the nanograting structures formed by DLW inside glass with a periodicity dependent on the pulse energy, which are accredited to a coupling between the electric field of the fabrication laser and excited electron plasma at the focus~\cite{kazansky03}.

Using the EELS analysis presented in Fig.~\ref{Fig3} (e), we estimate that only $\approx 4\%$ of the wire cross-sectional area contains conductive $sp^2$ bonded carbon. Indeed, even if it is assumed that all the regions identified as $sp^2$ are continuous along the wire and can hence contribute to any DC conductivity, values for the intrinsic structural resistivity are expected to be over an order of magnitude lower than those previously reported for laser written wires~\cite{Lagomarsino2013a, Sun2014b, Lagomarsino2014d, Shimizu2009, Trucchi2017}. The relatively low $sp^2$ content of even the patches in Fig.~\ref{Fig3}(e) is surprising and at no point are purely graphenic EELS spectra recorded. 

The work here clarifies a recent SEM analysis of subsurface DLW modifications in diamond~\cite{kononenko2016}, which also found only a small volume fraction of $sp^2$ bonded carbon in the laser irradiated zones. Here, the multiplicity of conductive regions and their spatial locations are revealed in detail. TEM images indicate the formation of discrete, 100~nm elliptical nanowires that are partially converted to $sp^2$ carbon. Stress-relieving dislocations radiate out from each region along $\left\{111\right\}$ planes and also contain trace quantities of $sp^2$ carbon. It is anticipated that both of these features contribute to electrical conductivity. 

In conclusion, the internal structure is revealed for laser written graphitic wires buried inside the bulk of diamond. While viewed in an optical microscope, the wires appear to show a bulk change from diamond to graphite, but the higher resolution of TEM shows that the structural modification is relatively sparse, comprising micro-cracks and nano-clusters of $sp^2$ bonded carbon. This disparity is particularly important when using the measured wire resistivity as a proxy measurement of wire composition. Indeed, the small effective wire cross-sections discovered here indicate the presence of a remarkably conductive carbonaceous phase. 

The authors gratefully acknowledge the Leverhulme Trust (RPG-2013-044) and the UK Engineering and Physical Sciences Research Council (EP/K034480/1) for financial support. We would additionally like to thank Philip Martineau and David Fisher for enlightening discussions.

%

\end{document}